\documentclass[
]{ceurart}

\usepackage{listings}
\usepackage{todonotes}
\usepackage{float}
\usepackage{rotating}

\lstdefinelanguage{json}{
  basicstyle=\ttfamily\footnotesize,
  showstringspaces=false,
  breaklines=true,
  keywords={id, golden, generated, order_matters},
  keywordstyle=\bfseries,
}

\begin{document}


\copyrightyear{}
% \copyrightyear{2026}
% \copyrightclause{Copyright for this paper by its authors.
%   Use permitted under Creative Commons License Attribution 4.0
%   International (CC BY 4.0).}

% \conference{ELMKE: Evaluation of Language Models in Knowledge Engineering 3rd Workshop co-located with ESWC 2026, Dubrovnik, Croatia}
\conference{}

\newcommand{\resourceName}{\texttt{t2s-metrics}\xspace}
\newcommand{\irMetrics}{\texttt{ir-metrics}\xspace}

\title{T2S-Metrics: Unified Library for Evaluating SPARQL Queries Generated From Natural Language}
\author[1,3]{Yousouf Taghzouti}[%
orcid=0000-0003-4509-9537,
email=yousouf.taghzouti@univ-cotedazur.fr,
url=https://youctagh.github.io/,
]
\author[1]{Tao Jiang}[%
orcid=0000-0002-5293-3916,
email=tao.jiang@cnrs.fr,
url=https://orcid.org/0000-0002-5293-3916
]
\address[1]{Univ. Côte d’Azur, CNRS, ICN, France}

\author[2]{Camille Juigné}[
orcid = 0000-0003-1157-9030,
email = camille.juigne@inria.fr,
url = https://cjuigne.github.io/cjuigne/
]

\author[1,3]{Benjamin Navet}[
orcid = 0009-0005-8523-0941,
email = benjamin.navet@univ-cotedazur.fr,
url = https://github.com/BenjaminNavet
]

\author[2]{Fabien Gandon}[%
orcid=0000-0003-0543-1232,
email=fabien.gandon@inria.fr,
url=http://fabien.info/,
]

\address[2]{Univ. Côte d'Azur, Inria, CNRS, I3S, France}

\author[2]{Franck Michel}[%
orcid=0000-0001-9064-0463,
email=franck.michel@inria.fr,
url=https://w3id.org/people/franckmichel,
]
\address[3]{Univ. Côte d'Azur, CNRS, Inria, I3S, France}

\author[1,3]{Louis-Felix Nothias}[%
orcid=0000-0001-6711-6719,
email=louis-felix.nothias@cnrs.fr,
url=https://lfnothias.github.io/,
]

\begin{abstract}
The evaluation of Question Answering (QA) systems over Knowledge Graphs has historically suffered from fragmentation, inconsistency, and limited reproducibility. While significant progress has been made in semantic parsing and SPARQL query generation, evaluation methodologies remain diverse, ad hoc, and often incomparable across studies. Existing benchmarks typically focus on a small subset of metrics, such as query exact match or answer-level F1, neglecting syntactic validity, semantic faithfulness, execution correctness, results ranking quality, and computational efficiency. 
In this paper, we present \resourceName, an open-source, extensible, and unified evaluation library designed specifically for SPARQL query comparison and execution-based assessment. \resourceName provides a broad and extensible set of over 20 evaluation metrics, collected from the literature and practical evaluation needs, spanning lexical, syntactic, semantic, structural, execution-based and ranking-based dimensions.
These include query-based metrics such as token-level Precision, Recall, and F1; BLEU, ROUGE, METEOR, and CodeBLEU variants; variable-normalized metrics (SP-BLEU, SP-F1); graph- and URI-based exact match metrics; as well as answer set–based metrics such as F1-QALD and Jaccard similarity; ranking metrics including MRR, NDCG, P@k, and Hit@k; and LLM-as-a-Judge metrics.
Taking inspiration from the \irMetrics library for Information Retrieval, \resourceName provides a modular abstraction layer that decouples metric specification from implementation, enabling consistent, transparent, and reproducible evaluation of SPARQL-based QA systems. We argue that \resourceName constitutes a necessary step toward systematic, standardized evaluation in question answering over knowledge graphs and facilitates deeper diagnostic insights into system behavior beyond answer correctness. % alone.
\end{abstract}


\maketitle

\section{Introduction}

The translation of natural language questions into formal query languages like SPARQL is a critical component of semantic search and Knowledge Graph Question Answering (KGQA)~\cite{Usbeck2019}. As models evolve from template-based systems to end-to-end neural networks and Large Language Models (LLMs), the requirements for evaluation have become increasingly complex~\cite{Papa2023}. Yet, current evaluation practices %in the KGQA community
are often fragmented. While one study might focus on surface-form metrics like BLEU~\cite{hamada2024} or ROUGE~\cite{Xueli2025}, effectively treating the SPARQL query generation as a machine translation task~\cite{Tommaso2018}, another might focus strictly on execution accuracy, that is, whether the query retrieves the correct answer~\cite{Samuel2023}, while ignoring the syntactic and semantic correctness of the generated query. A third might employ graph isomorphism to check for structural equivalence of two queries~\cite{Yu2020}. 

This fragmentation makes it difficult to compare results across papers, or to diagnose specific flaws in a model e.g., a model that generates syntactically correct SPARQL but queries non-existing graph patterns. Furthermore, the rise of generative AI has introduced new failure modes, such as "hallucination"~\cite{Aditya2025} where a model makes up plausible-looking but non-existing Uniform Resource Identifiers (URIs). 

Traditional metrics like lexical exact match or lexical similarity~\cite{Levenshtein1966} fail to capture the nuance of two queries that are semantically correct but syntactically distinct due to variations in variable naming or prefix definitions.
By contrast, the Information Retrieval (IR) community has long benefited from mature, standardized evaluation libraries~\cite{harman2011}. Tools such as \texttt{trec\_eval}~\cite{VanGysel2018pytreceval} and, more recently, \irMetrics~\cite{MacAvaney2022}, provide researchers with consistent access to a wide array of metrics, facilitating reproducibility and fair comparison across systems. Inspired by this paradigm, we argue that KGQA, and SPARQL query generation in particular, requires an equally systematic evaluation framework.
To this end, we introduce \resourceName (where \texttt{t2s} stands for text-to-SPARQL), a unified library for evaluating SPARQL queries generated from natural language, designed with four primary goals:

\begin{enumerate}
    \item \textbf{Comprehensiveness:} Support a broad taxonomy of metrics covering lexical similarity, structural equivalence, semantic correctness, execution validity, results ranking quality.
    \item \textbf{Abstraction:} Decouple metric specification from implementation details, enabling consistent evaluation across datasets and systems.
    %\fm{"metric definition" qu'est-ce que ça veut dire concretement ? Ailleurs on parle de metric specification. On a vraiment une séparation nette entre une spéficication (qu'est-ce que c'est) et d'une implémentation de chaque mesure ?}
    % On a la specification de la metric de base, ensuite d'une metric d'execution basée sur la première, et ensuite des metric qui implémentent ces interfaces.
    \item \textbf{Diagnosability:} Provide fine-grained signals that allow researchers to understand why a system succeeds or fails with specific metrics, rather than merely whether it does.
    \item  \textbf{Extensibility:} Provide researchers with a framework wherein they can easily add their own metrics and compare them to others.
\end{enumerate}

This paper presents the design principles, metrics taxonomy, and evaluation philosophy underpinning \resourceName, positioning it as a foundational tool for future KGQA research concerned with evaluation methods, benchmarks, and challenges. More specifically, this paper makes four contributions: (i) a taxonomy of evaluation dimensions for text-to-SPARQL systems, spanning lexical, structural, execution-based, ranking-based, and qualitative assessment perspectives;
%\fm{Dans l'abstract on parle de: lexical, syntactic, semantic, structural, execution-based, ranking-based. 
%structural = semantic? qualitative fait réf à quelles mesures? A uniformiser \textbf{sans blabla}.} 
% For the moment, in Structural, there is URI hallucination, which is neither lexical nor execution-based.
(ii) a unified evaluation framework that separates metric specification from implementation; (iii) an extensible open-source library, \resourceName, that operationalizes this framework; and (iv) an empirical illustration showing how multi-metric evaluation provides more informative diagnostics than single-metric assessment alone.

% ==================================================================================

\section{Background and Related Works}

\subsection{Text-to-SPARQL vs IR Evaluation}

The evaluation of text-to-code generation sits at the intersection of Natural Language Processing (NLP) and IR~\cite{harman2011}. However, evaluating SPARQL-generating QA systems presents challenges that are fundamentally different from traditional IR or text generation tasks:

\begin{itemize}
    \item \textbf{Structural sensitivity:} Minor changes in query structure (e.g., variable placement, OPTIONAL clauses) can drastically affect results.
    \item \textbf{Semantic equivalence:} Two SPARQL queries may be logically equivalent yet lexically dissimilar.
    \item \textbf{Execution dependency:} Query results correctness cannot be assessed independently of the underlying knowledge base, which entails reproducibility issues.
    \item \textbf{Partial correctness:} Queries may retrieve a subset of correct answers, over-generate results, or fail silently (e.g. returning an empty result without raising an error). %\fmt{\textit{[Je ne comprends pas le dernier point "fail silently"]}} \ytt{\textit{[Par exemple, sur Corese, si tu exécutes une requête qui utilise des préfixes sans les définir, tu obtiendras un retour vide, comme s'il n'y avait pas de résultat. C'est un exemple de Silent Failure.]}}
    \item \textbf{Ranking and ordering:} Some SPARQL queries produce ranked result sets, for which ranking-aware metrics are necessary.
\end{itemize}

\noindent Various families of metrics have been used over the years:
% \yt{
% Precision Token, Recall Token, F1 Token, Query execution, Macro-F1, F1-QALD, Precision of AnswerSet, , Recall of AnswerSet, runtime, Micro F1, Micro Precision, Micro Recall, Macro Precision, Macro Recall, F1 of AnswerSet \cite{Usbeck2019}. 
% BLEU-k, Cumulative BLEU \cite{Papineni2002}.
% Rouge-k\cite{Lin2004}.
% Precision@k, Recall@k, F1@k \cite{S2023}.
% SP-Bleu, SP-F1 \cite{Md.2022}.
% Levenshtein \cite{Levenshtein1966}.
% Jaccard \cite{Jaccard1901}.
% Cosine sim, P@1, MRR, Hit@k,\cite{Manning2008}.
% F1-SPINACH, EM SPINACH \cite{Shicheng2024}.
% METEOR \cite{Banerjee2005}.
% CodeBleu \cite{Ren2020}.
% URI EM, URI Hallucination \cite{Aditya2025}.
% Relaxed Exact Match (success), Relaxed Exact Match (all) \cite{Xueli2025}.
% NDCG Normalized Discounted Cumulative Gain \cite{Jarvelin2002}.
% sparqlIris, sparqlIriSuffix, $0.2 \times parse + 0.8 \times f1$ \cite{LarsPeter2024}.
% Pass@k \cite{Chen2021}.
% Success - 118 \cite{Vincent2024}.
% Granularity \cite{Kilian2025}.
% Determinism Score \cite{Alessandro2026}.
% sparqlIrisF1measure \cite{Desiree2025}.
% }

\begin{itemize}
    \item \textbf{Lexical Overlap Metrics:} Traditionally, systems have been evaluated using metrics borrowed from Machine Translation, such as BLEU~\cite{Papineni2002} and ROUGE~\cite{Lin2004}, also called surface form metrics. While useful for measuring n-gram overlap, these metrics often penalize valid SPARQL queries that use different variable or prefix names, different triple pattern ordering strategies, or semantically equivalent writing alternatives, e.g. using either the FILTER vs. VALUES SPARQL clauses.
    \item \textbf{Execution-Based Metrics:} To address the aforementioned limitations of surface forms, researchers utilize execution accuracy (Precision/Recall of the answer set)~\cite{S2023}. However, this requires a live endpoint with ''frozen'' dataset, and does not account for queries that are valid but return empty sets intentionally, nor does it penalize inefficient query structures.
    \item \textbf{Semantic and Structural Normalization:} Recent work has highlighted the need for query normalization, renaming variables and expanding prefixes, before comparison. Metrics like SP-BLEU~\cite{Md.2022} attempt to bridge the gap between surface form and logical structure.
\end{itemize}

\noindent Consequently, no single metric is sufficient to characterize the performance of a system~\cite{Dividino2013}.

\subsection{Libraries and Tools for Evaluating Text-to-SPARQL Performance}
This section overviews libraries and benchmarks relevant to measuring the performance of text-to-SPARQL systems. Where applicable, we note whether the tool provides only a subset of evaluation metrics or lacks direct support for SPARQL query evaluation.

\begin{enumerate}

\item \textbf{LMs4Text2SPARQL} A codebase for training and evaluating language models on text-to-SPARQL tasks. It includes scripts to compute standard evaluation metrics such as exact match and query execution accuracy.%
  \footnote{\url{https://github.com/AKSW/LMs4Text2SPARQL/blob/main/eval.py}}

\item \textbf{Text2SPARQL} A repository implementing grammar-pretraining and evaluation pipelines for models that output SPARQL. Includes metrics for SPARQL validity and execution correctness.%
  \footnote{\url{https://github.com/cskyan/Text2SPARQL/blob/master/utils/spider_evaluation.py}}

\item \textbf{LLMs-for-SPARQL} A collection of datasets and evaluation scripts to compare multiple text-to-SPARQL models (e.g., benchmark results across LC-QuAD, Spider4SPARQL).%
  \footnote{\url{https://github.com/SandroGT/LLMs-for-SPARQL/blob/main/experiments/evaluation/metrics.py}}

\item \textbf{text2sparql-client} A Python client to interact with text-to-SPARQL-compliant web endpoints. Useful in evaluating endpoint responses, success rates, and basic execution behaviors, but \emph{not} a full evaluation framework for synthetic metrics like F1.%
  \footnote{\url{https://github.com/AKSW/text2sparql-client/blob/main/text2sparql_client/utils/evaluation_metrics.py}}

\item \textbf{liita-nl2sparql} A Python package that implements a limited set of evaluation metrics for text-to-SPARQL models (e.g., syntactic validity, execution success). It \emph{does not} provide a comprehensive suite of semantic parsing metrics out of the box.%
  \footnote{\url{https://github.com/tonazzog/nl2sparql/blob/main/nl2sparql/evaluation/evaluate.py}}

\item \textbf{ir\_measures}  A Python library for calculating classical information retrieval metrics such as precision, recall, and nDCG. It is \emph{not} designed for SPARQL evaluation and thus does not natively handle SPARQL query results or measures like query execution correctness.%
  \footnote{\url{https://github.com/terrierteam/ir_measures}}

\item \textbf{pytrec\_eval}  A Python toolkit for computing evaluation measures from ranked retrieval results (e.g., for TREC-style ad-hoc search). Like \texttt{ir\_measures}, it provides IR metrics but \emph{does not} support SPARQL or semantic parsing outputs directly.%
  \footnote{\url{https://github.com/cvangysel/pytrec_eval/tree/master/py}}

\end{enumerate}

% \fm{Ici on a fait un tour des métriques, mais il manque un point sur les libs et frameworks comparables à t2s-metrics. Dans l'intro on parle de \texttt{trec\_eval} et \irMetrics, ici il faut dire en quoi ils ne suffisent pas, ou ne sont pas adaptés au text-to-SPARQL, ce qui permet de  justifier qu'on propose un nouveau framework t2s-metrics.}

\resourceName builds upon these foundations by aggregating these diverse methodologies into a single library. It complements dataset providers like QALD or LC-QuAD by focusing specifically on the computation of evaluation metrics.

% ==================================================================================

\section{The \resourceName Library}
The \resourceName Python library provides access to a broad and extensible suite of evaluation metrics tailored to SPARQL generation and KGQA. Table~\ref{tab:list-metrics} provides a summary of the supported metrics, categorized by their evaluation focus: Lexical, Semantic/Structural, Execution, and Model Performance. A column is dedicated to exemplifying the usage of the metric in literature. 
Note that Ollama\footnote{\url{https://ollama.com/}} is supported out of the box for local evaluation of the LLM-as-a-Judge metric, with a system evaluator prompt that can be edited if needed\footnote{\url{https://github.com/Wimmics/t2s-metrics/blob/main/t2smetrics/llm/ollama\_backend.py}}. By default the system prompts the LLM to act as a SPARQL expert, scoring a given query from 0.0 to 1.0 based on correctness, efficiency, readability, and adherence to best practices. The LLM returns a structured JSON response containing both a numerical score and a textual reasoning explanation. The method parses this response, handles errors gracefully (defaulting to a score of 0 if parsing fails), and returns a dictionary with the score, reasoning, and raw output.

% \yt{For cosine similarity, it is used usually in T2S systems to measure the similarity between queries as we did in Q²Forge / MetaboT / ChatExpasy. Should we add it to the list? } 
% \fab{what would be a reason not to?}
% \yt{There is no reason not to. It's just that the other metrics were used in the evaluation of golden/generated, rather than for comparing queries as part of the pipeline}
% \fab{there is no harm in adding it then}

\begin{sidewaystable}
\centering

% \fab{two references missing in the table}

% \begin{table}[]
\begin{tabular}{llp{16cm}l}
\textbf{Category}   & \textbf{Metric}                   & \textbf{Description}                                                                 & \textbf{Usage} \\
Lexical    & BLEU@K / BLEU C          & Measures k-gram precision against reference. Supports cumulative k-gram aggregation (Geometric mean) & \cite{hamada2024} \\
\textbf{}  & SP-BLEU                  & Adaptation where variables are normalized before standard BLEU calculation                   & \cite{Md.2022} \\
\textbf{}  & CodeBLEU                 & Evaluates code synthesis quality, incorporating syntactic AST information                           & \cite{Mehdi2025} \\
\textbf{}  & Cosine Similarity        & Calculates the cosine similarity score between the queries by converting them into numerical vectors using a bag-of-words approach & \cite{Yousouf2025} \\
% \textbf{}  & Euclidean Distance       & the similarity between two queries is calculated by first transforming them into high-dimensional binary vectors and then measuring the distance between those vectors        & \cite{TBD} \\
\textbf{}  & Exact Match (Token)              & Binary metric (1 or 0) indicating if prediction is identical to reference.                           & \cite{Aditya2025} \\
\textbf{}  & F1 (Token)               & The harmonic mean of Precision and Recall (AnswerSet)                          & \cite{Papa2023} \\
\textbf{}  & SP-F1                    & Adaptation where variables are normalized before standard F1 (Token) calculation & \cite{Md.2022} \\
\textbf{}  & Jaccard                  & Intersection over Union of token sets between generated and target queries.                          & \cite{Mehdi2025} \\
% \textbf{}  & Levenshtein              & Minimum single-character edits (insert, delete, sub) to transform prediction to reference.           & \cite{TBD} \\
\textbf{}  & METEOR                   & Matches unigrams based on exact match, stemming, and synonymy.                                       & \cite{Jens2023} \\
\textbf{}  & Precision/Recall (Token) & Proportion of correct answers retrieved vs. total retrieved (Precision) or total gold (Recall).      & \cite{hamada2024} \\
\textbf{}  & ROUGE@K                  & Assesses overlap of unigrams and bigrams, ect. between model generation and ground truth            & \cite{Xueli2025} \\
Structural & URI Hallucination        & Percentage of queries containing URIs that do not exist in the KB memory.                            & \cite{Aditya2025} \\
% \textbf{}  & EM Isomorphism         & Exact match based on graph isomorphism of the parsed SPARQL algebra.                                 & \cite{TBD} \\
% \textbf{}  & Syntax Score           & Ratio of syntactically valid queries (parseable) to total generated queries.                         & \cite{TBD} \\
% \textbf{}  & URI EM                 & Evaluates if the set of URIs in the prediction matches exactly with the reference.                   & \cite{TBD} \\
Execution  & F1 (AnswerSet)           & The harmonic mean of Precision and Recall (AnswerSet)                          & \cite{Samuel2023} \\
\textbf{}  & F1-QALD                  & Specialized F1 for Linked Data; handles empty set comparisons specifically                          & \cite{hamada2024} \\
\textbf{}  & F1-Spinach               & Specialized F1 for Linked Data; permits additional columns in the predicted results without penalty & \cite{Shicheng2024} \\
\textbf{}  & SP-F1                    & Adaptation where variables are normalized before standard F1 calculation & \cite{Md.2022} \\
\textbf{}  & Query Execution          & Measures successful endpoint completion (no errors), acknowledging valid empty sets.                 & \cite{hamada2024} \\
\textbf{}  & Precision/Recall (AnswerSet) & Proportion of correct answers retrieved vs. total retrieved (Precision) or total gold (Recall).      & \cite{Ruijie2019} \\
% \textbf{}  & Relaxed EM                & Success rate considering only syntactically valid and non-empty result queries.                      & \cite{TBD} \\
Ranking    & Hit@K                    & Measures whether at least one relevant item appears in the top K results                                               & \cite{Jens2023} \\
\textbf{}  & MRR                      & Evaluates the position of the correct answer in a ranked list            & \cite{Reham2025} \\
\textbf{}  & NDCG                     & Normalized Discounted Cumulative Gain; penalizes correct answers appearing lower in rank.            & \cite{ReneNone} \\
\textbf{}  & Precision@k              & Measures whether the top K results are relevant            & \cite{S.2023} \\
\textbf{}  & LLM-as-a-judge           & Prompts an LLM to qualitatively judge the query correctness given the question.     & \cite{bekbergenova2025metabot}
\end{tabular}
\caption{Metrics provided by \resourceName with a paper example using it.}
% \end{table}

\label{tab:list-metrics}
\end{sidewaystable}

\section{Interfaces and Implementation}
\resourceName is implemented in Python and designed to be easily integrated into training loops or post-hoc evaluation pipelines.
The library's \textbf{Python API} provides users with a natural syntax to compute metrics,
allowing for the seamless calculation of multiple metrics in a single pass. 
For example, computing SP-BLEU or F1-QALD requires only passing the predicted and gold-standard SPARQL queries, \resourceName handles the necessary pre-processing such as IRI normalization or variable renaming, automatically.

\subsection{Usage}
The example in Listing~\ref{listing:usage-example} demonstrates evaluating a set of predicted SPARQL queries against ground truth queries, requesting several lexical and execution-based metrics. The predicted and ground truth SPARQL queries are given in a file formatted as shown in Listing~\ref{listing:dataset-example}.

\begin{lstlisting}[basicstyle=\ttfamily\footnotesize, numbers=left,  breaklines=true, xleftmargin=20pt, xrightmargin=20pt, language=Python, caption=How to use \resourceName using the Python API, label=listing:usage-example]
from t2smetrics import run_experiments
from t2smetrics.metrics import ( LevenshteinDistance, AnswerSetF1, AnswerSetPrecision, 
  AnswerSetRecall, Bleu, CodeBLEU, SPF1, CosineSimilarity, EuclideanDistance, F1Spinach,
  HitAtK, JaccardSimilarity, Meteor, F1QALD, MRR, NDCG, SPF1, RougeN, PrecisionAtK,
  PrecisionQALD, QueryExactMatch, RecallQALD, QueryExecution, LLMJudge, SPBleu, TokenF1,
  TokenPrecision, TokenRecall, URIHallucination )

dataset_name = "example"
graph_path = "./datasets/example/kg/example.ttl"
jsonl_paths = ["./datasets/example/eval/example.jsonl"]
metrics = [
  AnswerSetPrecision(), AnswerSetRecall(), AnswerSetF1(), Bleu(), SPBleu(), CodeBLEU(),
  CosineSimilarity(), EuclideanDistance(), F1QALD(), PrecisionQALD(), RecallQALD(),
  F1Spinach(), HitAtK(k=5), JaccardSimilarity(), LLMJudge(), LevenshteinDistance(),
  MRR(), Meteor(), NDCG(), PrecisionAtK(k=1), QueryExecution(), QueryExactMatch(),
  RougeN(1), RougeN(2), RougeN(3), RougeN(4), TokenF1(), SPF1(), TokenPrecision(),
  TokenRecall(), URIHallucination()
]

run_experiments.run( dataset=dataset_name, jsonl_evals=jsonl_paths, metrics_list=metrics,
  execution_backend_graph_path=graph_path, verbose=True )
\end{lstlisting}

\begin{lstlisting}[basicstyle=\ttfamily\footnotesize, numbers=left,  breaklines=true, xleftmargin=20pt, xrightmargin=20pt, language=json, caption=An example of the file containing the predicted and ground truth SPARQL queries used by \resourceName, label=listing:dataset-example]
{"id": "q1", "golden": "SELECT ?x WHERE {?x a <http://example.org/Person>}", "generated": "SELECT ?y WHERE {?y a <http://example.org/Person>}", "order_matters": false}
{"id": "q2", "golden": "SELECT ?x WHERE {?x <http://example.org/age>30 }", "generated": "SELECT ?x WHERE {?x <http://example.org/age> 40}", "order_matters": true}
\end{lstlisting}

\textbf{Normalization Engine.} A core component of \resourceName is the normalization engine. When metrics like SP-BLEU and SP-F1 are requested, the engine automatically:
\begin{enumerate}
    \item Parses the SPARQL string;
    \item Resolves prefixes using a built-in or provided prefix map;
    \item Performs variable renaming normalization;
    \item Re-serializes the query for comparison.
\end{enumerate}

\subsection{Extending \resourceName}
The library can be easily extended with additional metrics. Developers only need to extend the base \texttt{Measure} class, or one of the more specialized variants (e.g., \texttt{AnswerSetMeasure}), and provide an implementation of the \texttt{compute} method, as illustrated in Listing~\ref{listing:extension-example}.

\begin{lstlisting}[basicstyle=\ttfamily\footnotesize, numbers=left,  breaklines=true, xleftmargin=20pt, xrightmargin=20pt, language=Python, caption=How to extend \resourceName with new metrics, label=listing:extension-example]
class MRR(AnswerSetMeasure):
    name = "mrr"

    def compute(self, case, context: EvaluationContext = None):
        gold, pred = self._get_answer_lists(case, context)

        if not self._validate(gold, pred):
            return EvaluationResult(case.id, self.name, 0.0)

        for i, ans in enumerate(pred):
            if ans in gold:
                return EvaluationResult(case.id, self.name, 1.0 / (i + 1))

        return EvaluationResult(case.id, self.name, 0.0)
\end{lstlisting}

\subsection{Source Code Availability} 
\resourceName is provided under the GNU Affero General Public License v3.0 or later (AGPL-3.0-or-later) license. The code is published on a public Github repository, and the version used at the time of writing us identified by a DOI to ensure the long-term preservation and citability. The library is published in Pypi. Table~\ref{tab:recap-links} summarises the links to the source code, demonstration videos and online prototype of \resourceName.

% \vspace{-15pt} % to eliminate gaps
\begin{table}[h!]
\centering
% {\normalsize % change table font size
\resizebox{\linewidth}{!}{  % Resize to fit page width
\begin{tabular}{lll}
%\hline
%\textbf{Resource type}                        & Reusable Software and Services \\ \hline
\textbf{License}                              & GNU Affero General Public License v3.0 or later (AGPL-3.0-or-later) \\ \hline
%\textbf{Online prototype}                     & \url{https://www.w3id.org/t2s-metrics/} \\ \hline
\textbf{PyPI package}                     & \url{https://pypi.org/project/t2s-metrics/} \\ \hline
\textbf{Repo \& DOI} & \url{https://github.com/Wimmics/t2s-metrics} - \href{https://doi.org/10.5281/zenodo.19255557}{10.5281/zenodo.19255557}         \\ \hline
% \textbf{Repo} & \url{https://github.com/Wimmics/t2s-metrics}  \\ \hline
\textbf{Demo video} & \textbf{Teaser} \url{https://youtu.be/w5nm7Rg5EeM} \textbf{Full} \url{https://youtu.be/ETmzMsnYEqY} \\ \hline
\end{tabular}%
}
\vspace{0.25cm} % to eliminate gaps
\caption{License and links to the source code, PyPI package, demonstration videos and online prototype of \resourceName.}
\label{tab:recap-links}
% }
\end{table}
% \vspace{-25pt} % to eliminate gaps

% ==================================================================================

\section{Practical Use Case and Evaluation}

\resourceName serves two primary user groups:
\begin{enumerate}
    \item \textbf{Model Developers} who need granular feedback. For example, a developer might see high BLEU but low URI EM (Exact Match), indicating their model understands SPARQL syntax but is memorizing/hallucinating incorrect entity identifiers.
    \item \textbf{Benchmark Creators} who need standard metrics to ensure fair comparison. Using \resourceName, new datasets can provide baseline scores across 20+ metrics.%, rather than just one.
\end{enumerate}

\begin{figure}[ht]
    \centering
    \includegraphics[width=\linewidth]{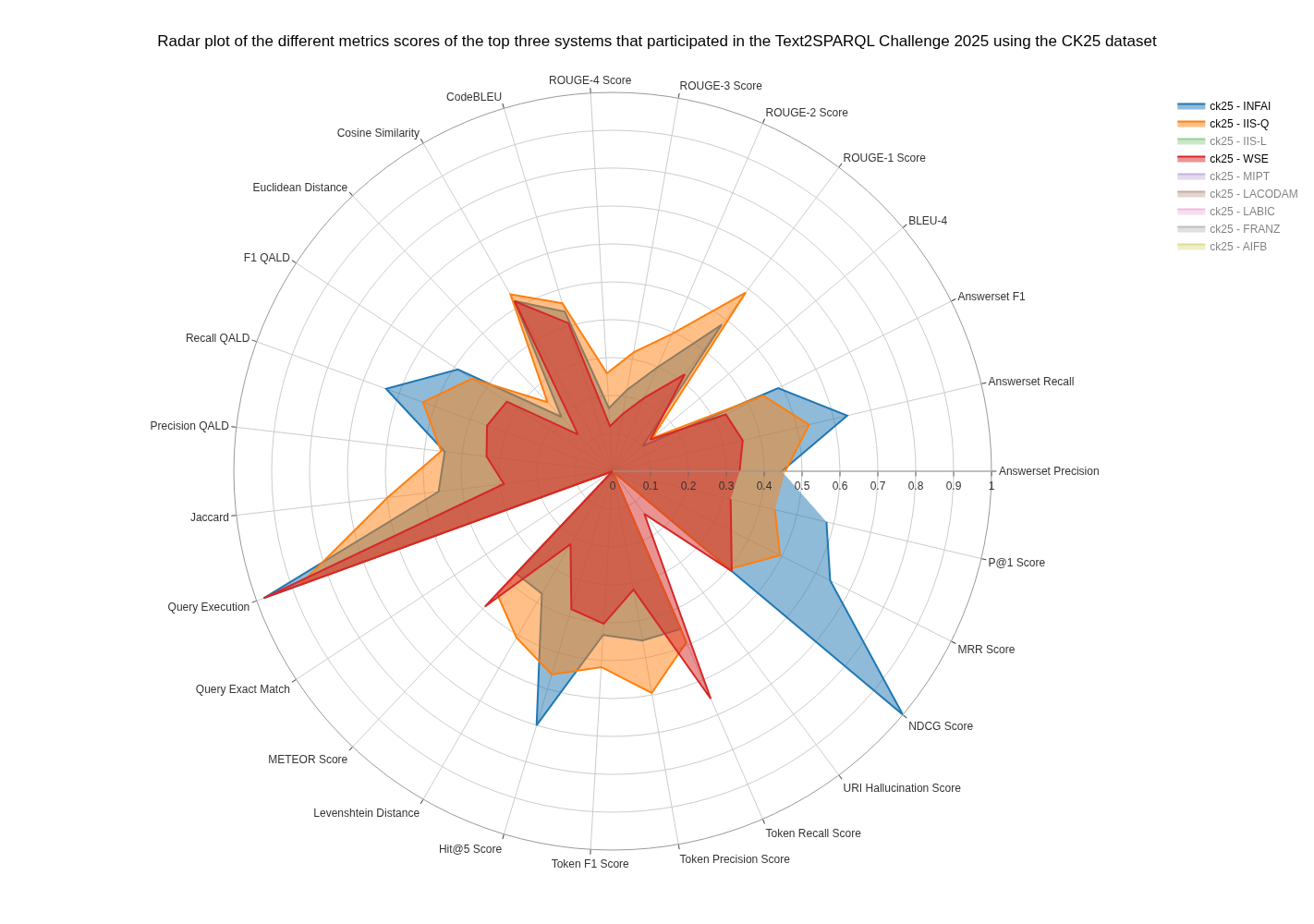}
    \caption{Radar plot of the different metrics scores of the top three systems that participated in the Text2SPARQL Challenge 2025 using the CK25 dataset.}
    \label{fig:radar_top3}
\end{figure}

To demonstrate \resourceName, we used the results produced by the KGQA systems that participated in the Text2SPARQL 2025 challenge,\footnote{\url{https://text2sparql.aksw.org/2025/results/}} and evaluated them using the metrics we provide on the \textit{ck25} dataset. 

\begin{figure}[ht]
    \centering
    \includegraphics[width=\linewidth]{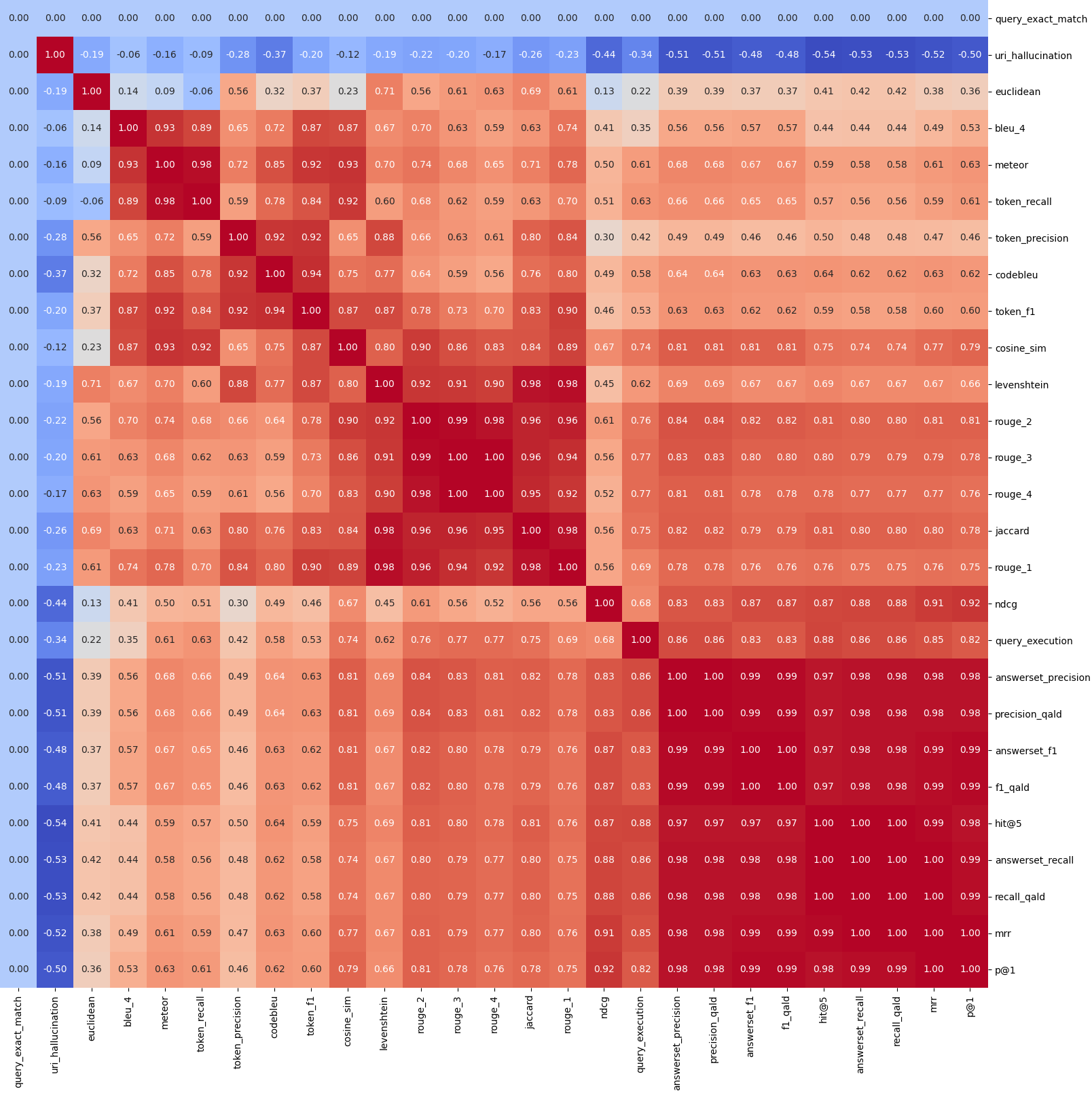}
    \caption{Correlation matrix between the different metrics applied to the systems that participated in the Text2SPARQL Challenge 2025, using the CK25 dataset.}
    \label{fig:corr_matrix}
\end{figure}

Figure~\ref{fig:radar_top3} presents a radar plot showing the metric scores of the top three systems for the \textit{ck25} dataset, while Figure~\ref{fig:corr_matrix} displays a correlation matrix of the different metrics scores, automatically grouped according to their pairwise correlations.
From the correlation matrix, we observe a clear clustering pattern. The bottom-right section shows a concentration of execution-based metrics, whereas the middle section primarily contains string-based metrics (applied to the generated query). Notably, there is a strong negative correlation between the \textit{URI hallucination} metric and the execution-based metrics. This relationship is expected, as an increase in hallucinated URIs often prevents successful query execution, resulting in lower execution scores.

This negative correlation is less pronounced for string-based metrics, with the exception of the \textit{CodeBLEU} metric. This can be explained by the fact that \textit{CodeBLEU} evaluates syntactic correctness by parsing the query, making it more sensitive to errors caused by hallucinated URIs. Since the systems participating in the challenge relied on language models to generate queries, none of their queries matched the gold queries exactly, which explains the zero correlation scores with the \textit{query exact-match} metric.

% ==================================================================================

% ==================================================================================

\section{Conclusion}

We presented \resourceName, a unified evaluation toolkit for SPARQL generation and KGQA. 
By aggregating over 20 metrics, including recent adaptations like SP-BLEU and URI Hallucination, \resourceName addresses the fragmentation in current evaluation methodologies. 
It allows researchers to move beyond simple text matching and rigorously assess the syntactic validity, semantic accuracy, and execution reliability of their systems. We hope \resourceName will become a standard utility in the semantic web community, fostering more transparent and reproducible evaluation. We acknowledge that \resourceName has limitations in measuring certain operational metrics, including LLM input/output tokens, token pricing, and GPU usage, as these are inherently difficult to compute without access to the internal mechanisms of the underlying methods. As future work, we plan to develop \resourceName hooks that can be integrated into the implementation to enable reliable capture of these metrics.

\begin{acknowledgments}
This work was supported by the French Government's France 2030 investment plan (ANR-22-CPJ2-0048-01), through 3IA Cote d'Azur (ANR-23-IACL-0001) as well as the MetaboLinkAI bilateral project (ANR-24-CE93-0012-01 and SNSF 10002786). Additional support was provided through the UCAJEDI Investments in the Future project (ANR-15-IDEX-01).
\end{acknowledgments}

%% The declaration on generative AI comes in effect
%% https://ceur-ws.org/GenAI/Policy.html
\section*{Declaration on Generative AI}
During the preparation of this work, the author(s) used ChatGPT, Gemini and DeepL for the following: drafting content, grammar and spelling checks.
After using these tools/services, the author(s) reviewed and edited the content as needed, taking full responsibility for the publication's content.

%%
%% Define the bibliography file to be used
\bibliography{sample-ceur}

\end{document}